# Symmetry engineering in 2D bioelectronics facilitating augmented biosensing interfaces

Yizhang Wu[1+], Yihan Liu[1+], Yuan Li[2+], Ziquan Wei[3], Sicheng Xing[1], Yunlang Wang[4], Dashuai Zhu[5], Ziheng Guo[1], Anran Zhang[1], Gongkai Yuan[1], Zhibo Zhang[6], Ke Huang[7], Yong Wang[8]*, Guorong Wu[3,9,10,11]*, Ke Cheng[5], and Wubin Bai[1]*

[1]Department of Applied Physical Sciences, University of North Carolina, Chapel Hill, NC 27599, USA

[2]Joint Department of Biomedical Engineering, North Carolina State University; University of North Carolina at Chapel Hill, Raleigh, NC, 27607, USA

[3]Department of Computer Science, University of North Carolina at Chapel Hill, Chapel Hill, NC, 27599, USA

[4]Department of Physics, Nanjing University, Nanjing, Jiangsu, 210000, China

[5]Department of Biomedical Engineering, Columbia University, New York, NY, 10027, USA

[6]Department of Computer Science and Engineering, Michigan State University, East Lansing, MI, 48824, USA

[7]Department of Molecular Biomedical Sciences, North Carolina State University, Raleigh, NC, 27607, USA

[8]Wide Bandgap Semiconductor Technology Disciplines State Key Laboratory, School of Microelectronics, Academy of Advanced Interdisciplinary Research, Xidian University, Xi'an 710071, China

[9]Department of Psychiatry, University of North Carolina at Chapel Hill, Chapel Hill, NC, 27599, USA

[10]Department of Statistics and Operations Research, University of North Carolina at Chapel Hill, Chapel Hill, NC, 27599, USA

[11]UNC Neuroscience Center, University of North Carolina at Chapel Hill, Chapel Hill, NC, 27599, USA

[+]Contribution equally

*Corresponding author:

yongwang@xidian.edu.cn

guorong_wu@med.unc.edu

wbai@unc.edu

**Abstract**:

Symmetry lies at the heart of 2D bioelectronics, determining material properties at the fundamental level. Breaking the symmetry allows emergent functionalities and effects. However, symmetry modulation in 2D bioelectronics and the resultant applications have been largely overlooked. Here we devise an oxidized architectural MXene, referred as OXene, that couples orbit symmetric breaking with inverse symmetric breaking to entitle the optimized interfacial impedance and Schottky-induced piezoelectric effects. The resulting OXene validates applications ranging from microelectrode arrays, gait analysis, active transistor matrix, and wireless signaling transmission, which enables highly-fidelity signal transmission and reconfigurable logic gates. Further OXene interfaces are investigated in both rodent and porcine myocardium, featuring high-quality and spatiotemporally resolved physiological recordings, while accurate differentiated predictions, enabled via various machine learning pipelines.

**Summarize in one sentence:**

Symmetry engineering in 2D bioelectronics implements highly-fidelity signal transmission and reconfigurable logic gates.

**Main:**

Bioelectronics with 2D materials enable paradigm-shifting diagnostic and therapeutic strategies, owing to a plethora of distinctive phenomena, such as rectification, the photovoltaic effect, and the quantum Hall effect(*1–3*). Of particular interest are these subjects that offer customized modes of operation enabled by these versatile components as electrical/optical interfaces across a range of spatiotemporal scales. Recent progress in such 2D bioelectronics establishes the basis for seamless integration of microsystems technologies with living organisms, to provide persistent, multimodal function with applications in modulating cardiac cycles, deep-brain stimulation, regenerating sensorimotor functions, and many others(*4–7*). Thus, interfacing 2D bioelectronics with biological tissues represents a promising trend in probing and actuating biological systems. The broad central focus in 2D bioelectronics that requires reliable interaction between biological tissues and electronics is in the development of augmented biosensing interfaces, underlying a broad collection of application scenarios, featuring high-quality physiological recording with chronical stability and biocompatibility(*8*).

In this regard, symmetry engineering in 2D bioelectronic materials holds great promise to provide intrinsic augmentation in targeted functionalities. Conceptually differing from functionalized device integration, symmetric breaking offers a distinctive approach to improving signaling transmission or energy transduction beyond the conventional limitation imposed by inherent material properties. For example, the piezoelectric effect, featuring non-centrosymmetric materials, converts mechanical energy into electricity, furthering untethered, self-powered electronics(*9, 10*). Nevertheless, material symmetry is generally determined by its pristine crystallographic structure, and loss of symmetry usually occurs *via* phase transitions(*11*); Or the symmetry can be tuned by external stimuli, such as the flexoelectric effect induced by strain gradients(*12*), while is hampered by its rather small effective coefficients and a complicated setup for inducing large strain gradients(*13*). Thus, an alternative, subjected to robust and spontaneous, would be highly desirable for developing 2D bioelectronic interfaces based on symmetry breaking, while connecting these fundamental insights to real-world applications, especially in clinical practice.

We devise an oxidized-MXene (OXene) architectural composite that emphasizes decoupling the broken symmetry configuration into orbit symmetric breaking and inverse symmetric breaking, which cooperatively augments electron transport and response (**Fig.1A**). Such augmentation implements the exploiting and coupling of additional out-of-plane electron conduction and built-in polar structures. The subsequently improved electronic-tissue impedance and Schottky-induced piezoelectric effect both enable the augmented biosensing interfaces via various device configurations (**Fig.1B**). This approach allows for extensive versatility for varying 2D bioelectronic applications, including bioelectrode array, gait analysis, wireless implants, and reconfigurable computational substrate in transistor, with capabilities of high-fidelity operation and scalable spatiotemporal resolution. Incorporated with machine learning (ML) pipelines, data obtained by OXene-infused interfaces can be used to classify physiological signals at high accuracies and provide reliable predictions on potential adverse events.

Synthesis of OXene leverages a facile strategy (Fig.S1), with three primary steps: (i) constructing intrinsically conductive traces by solution processable MXene, (ii) implementing on-demand transformation by laser cutting, (iii) deriving anatase $TiO_2$ on the surface by oxygen plasma treatment, collectively as the key innovation. Completely isolated symmetry breaking induction facilitates OXene with capabilities to realize local region customizations upon varying functional bioelectronics.

The controlled oxidization process of MXene may play a proximate role in the strain gradient in terms of symmetry engineering, lowering or even breaking the symmetry. Morphologies of OXene reveal spindle-shaped anatase $TiO_2$(*14*) immobilized at flake edges (Fig.S2a-d). OXene preserves the basic infrastructure of MXene, while reorganizing the surface dangling bonds and regulating the out-of-plane vibration (Fig.S3). The derived oxides, localizing charge in the form of polarons on metallic surface, can act as deep-state point defects, which render a charge transfer pathway differing from pristine planar electron transport (Fig.S2e). Besides, evidence that anatase $TiO_2$ sustains large electron polaron, which promotes the frontier orbitals (conduction-band minima, CBM) made up of weakly interacting orbitals and a high density of states around the band edges, favoring symmetry reduction(*15*). Theoretical calculations (**Fig.1C** and Fig.S4) predict that *p-d* hybridization alters orbital symmetry in OXene, advancing misalignment of metallic d orbitals with $t_{2g}$ symmetry and O *2p* orbitals(*16*). Besides, a dipole transition in OXene features Ti 1 s orbital to the p-component (Fig.S4d), specifically to projected $P_z$ orbital of oxygen, which is distinguished from other titanium species (e.g., compounds, monomers, and oxides) (**Fig.1D** and Fig.S5). Symmetry breaking in $P_z$ orbitals realigns out-of-plane vibrations in R-space (Fig.S5d), with capabilities of actively participating in electron transfer(*17*).

The out-of-plane electron transfer entitles to an improved interfacial impedance (**Fig.1E** and Fig.S6), compared with other common bioelectrical traces (e.g., Au, PEDOT:PSS, MXene). The derived $TiO_2$ facilitates a heterogeneous semiconductor-metal interface, while creating a depletion or space charge region. Double layer capacitance ($C_{dl}$, Fig.S6e) consequently forms in the presence of electrolytes, accompanied by anisotropic out-of-plane/in-plane electron conduction, which contributes predominately to reducing impedance(*18*). The charge transfer resistance ($R_{ct}$, Fig.S6g) in OXene, providing high charge-carrier mobility, optimizes its interfacial impedance as well.

The derived oxides coupled with metallic substrates, considering the varying terminal groups (-F, -OH, and =O), that enable to form metal–semiconductor contacts termed Schottky junctions (Fig.S7). Schottky junctions feature a rather strong built-in field, modulating the partially filled valence Ti *d*-orbitals (Fig.S8a), which allows the rearrangement of the energy levels for aligning the Fermi level (Fig.S8b). Schottky barrier renders charge accumulation in metallic substrates while a depletion region within the anatase $TiO_2$ (Fig.S8c)(*19*). The carrier tunneling behavior (Fig.S8d) also enhances the built-in electrical field and induces interface polarization(*20*). Massive charge transfer and charge redistribution lead to discontinuities in the interfacial potential and the presence of interface dipoles(*21*).

Schottky junction induces the piezoelectric effect in OXene, showing deformation aligned linearly with potential, whereas MXene barely exhibits a piezoelectric excitation (Fig.S9). After eliminating morphological effects employing dual frequency resonance tracking (DFRT), we achieve uniform and stable out-of-plane and in-plane piezoelectric responses (Fig.S10). Phase signals (Fig.S10b) illustrate a consistent polarization direction with electrical field, demonstrating the presence of piezoelectric dipoles in OXene(*22*). The piezoelectric response in OXene is aligned with the simple harmonic oscillator (SHO) model, conforming to out-of-plane (first harmonic) and in-plane (second harmonic) resonance peaks (**Fig.1F**). OXene shows a higher second harmonic response than the first harmonic response (Fig.S11), underlying piezoelectricity primarily arises from polar dipoles rather than spontaneous polarization or electrostatic interference(*23*).

Polarization-resolved Second Harmonic Generation (SHG) displays a tri-fold rotational symmetry dependent on the azimuthal angle of monolayer OXene (**Fig.1G**). The SH response stems from the broken inversion symmetry. Asymmetric engineering results in local polarization via the polar nature of the built-in field, emerging piezoelectric effects(*24*). Accordingly, dual symmetry breaking in OXene promises to facilitate a variety of practical bioelectronics with the fundamental principles.

**OXene-infused Bioelectrode Interfaces for pathological predication**

Improving interfacial impedance at the electronic-tissue interface potentially renders high-fidelity signaling transport and built-in signal amplification for implantable sensors. To demonstrate enhanced interfacial impedance enabled by OXene, we interfaced a 3 x 3 bioelectrodes array with the rat epicardium under sinus and myocardial infarction (MI) (**Fig.2A** and Fig.S12a). The multichannel electrode captures electrocardiography (ECG) covering the left ventricle (**Fig.2B**). Built on a flexible styrene-ethylene-butadiene-styrene (SEBS) substrate (~ 20 μm), this array conformally laminates on the epicardium against rhythmic cardiac systole and diastole (**Fig.2C,** video.S1 and Fig.S12b).

To comprehensively character the signal qualities, we compared unfiltered raw data using randomly mounted OXene and MXene bioelectrodes collected from various rodent models with Sinus or MI conditions (Fig.S12c). Myocardial mechanical motions can contaminate electrophysiological signals with unwanted noise. Improved impedance in OXene at electronic-tissue interface ensures superior signal fidelity and higher signal-to-noise ratio (SNR) compared with MXene, enabling more precise cardiac monitoring. Specifically, OXene-infused bioelectrodes allow for clearly delineating P-wave (represents atrial depolarization) and QRS complex (represents ventricular depolarization) in the acute infarction model (**Fig.2D**), which is characterized by slow ventricular depolarization and conduction system dysfunction(*25*). This clarity stems from improved interfacial impedance and is crucial for assessing patient cardiac status in clinical settings. Conversely, signals collected from MXene electrodes exhibit baseline fluctuations and inferior signal noise.

The high recording performance offered by the OXene-infused interfaces enable high-fidelity ML evaluations to predict pathological risks of cardiac tissues at early stage. The observation on the OXene signal distribution (**Fig.2F**) shows a separate cluster of cardiac diseases (e.g. MI), where each point that is a 9-channel vector is closer to each other under the same cardiac status in a 2-dimensional space, using the TSNE dimensionality reduction algorithm, which preserves the relationship between the original data points. To quantitively measure the benefit of OXene on Sinus/MI classification, we comprehensively test a deep neural network (DNN) (**Fig.2E**) that corporates with the 8-way-1-shot supervised learning, i.e., each class has one data point for training the DNN, in all possible cases. Accuracy comparison of precision-recall curves (**Fig.2G**) illustrates that OXene mostly outperforms MXene, which tends to be solely high on recall or precision. Additionally, six classes of OXene are more accurate than MXene on the confusion matrix (**Fig.2H** vs. Fig.S13a) where the deep learning task involves different locations of the sensor settled, 4 for Sinus/MI as mentioned above, leading to an 8-class classification. In case they are separately trained, 3 out of 4 locations show that OXene enables better performance on the confusion matrix (Fig.S13b-c). Such superiority of OXene to recognize cardiac disease is supported by the higher accuracy (0.9~1.0) on train and evaluation curves (Fig.S13d-g).

OXene array is capable of spatial and temporal mapping over large areas of the heart. Representative sinus and MI segments are intercepted, while selecting four R-wave (represents the maximum amplitude in QRS complex) timestamps (**Fig.2I**). Instantaneous ECG mapping can be interpolated across the matrix and normalized, illustrating the propagated spatial disparities in electrophysiological activity between the sinus and MI models over time. It is noted that the spatial resolution can be further extended with denser sensing nodes.

OXene array not only captures localized ECG fragments but also records long-term cardiac activity, featuring consistently high SNR and spatiotemporal resolution (Fig.S14). The expanded and dynamic ECG mapping (Video.S2) offers further insights into arrhythmia pathogenesis and can be synchronized with structural computed tomography images of the heart for precise arrhythmia focus localization. In this view, solution-processable OXene traces also approach them as excellent coatings on the electrode surface of existing commercially implantable electronics including cardiac pacemakers, neurostimulators, deep brain stimulators, and others, for augmented biosensing interfaces.

**Schottky interface from OXene-filled composite for Gait Analysis**

Piezoelectrically-mediated OXene allows extra polarized charges with external stress via electrostriction effect, promising to achieve impressive signal sensitivity in pressure sensing electronics.

To visualize and quantitatively characterize the piezoelectric effect, we designed an OXene-filled ionogel architecture anchored onto the interdigital electrodes for gait analysis (**Fig.3A**). The choice of the specific ionogel as pressure modules was motivated by its ionical conductivity, ultra-toughness, and stretchability(*26*). These ionogels render hydrogen-bonded domains connected by highly solvated and soft domains to form an extremely tough ionogel (**Fig.3B**), enabling stable and uniform dispersing of pressure from curvilinear surfaces (Fig.S15a). Whereas the high fracture strength (21.6 mPa) and

Young's modulus (35.9 mPa) of ionogels consequentially contribute to strain insensitivity and signal vulnerability (Fig.S15b-c).

OXene flakes form in-plane conductive percolation networks that are strongly crosslinked with hydrogen bonding employing electrostatic self-assembly. The anatase $TiO_2$ isolated from in-plane networks facilitates anisotropic out-of-plane electrical conduction (**Fig.3C**). Consequently, the built-in field-induced piezoelectric effect allows for charge redistribution at the Schottky interface under tensile or compressive forces (**Fig.3D**), resulting in interlayer displacement of charge carriers that correspondingly augment the sensitivity of the stress response.

Mechanistically, we used a dynamic theoretical calculation to intuitively simulate the band structure subjected to varying strains (Fig.S16-S18). The deformation facilitates a transition from P-type Schottky contact to N-type Schottky contact or N-type quasi-ohmic contact within OXene, redistributing the charge between the metal and the semiconductors to reach a new equilibrium state(*27*). Subsequently, the effective barrier of the Schottky junction, featuring insulating properties, prevents the polarized electrons from flowing across the interface directly, thus outputting electricity through the external circuit(*28*). Meanwhile, the built-in field under mechanical strain achieves enhanced electric polarization in the depletion region (Fig.S19), as a driver for efficient separation of electron-hole pairs(*29*). Among these, density of states of Ti atoms near the Fermi level dominates piezoelectric polarization process. It is worth noting that the basic Schottky model is used here for simplicity, without considering the electric-field-dependent permittivity and anomalous capacitance-voltage characteristics.

Compared with MXene-enriched ionogels, OXene-enriched ionogels indicate a markedly enhanced pressure response (**Fig.3E**). We then deployed a real-time control system on the subject including an OXene-infused multi-matrix sensor, clock signal generator, signal preprocessing circuit, and MCU with wireless transmitter module (**Fig.3F** and Fig.S20). Notably, adaptation of an analog-to-digital converter (ADC) allow sampling of the impedance magnitude of the composite, rather than the resistance, avoiding signal drift and inaccurate measurement caused by the viscoelastic creep of the soft materials (Fig.S21)(*30*). This sensory system allows for continuously and synchronously capturing the pressure distribution on the sensing pixels featuring drift-free signals under prolonged high stresses (**Fig.3G** and video.S3).

OXene-infused ionogel, featuring highly sensitive and stable sensing capabilities, offers superior recognition accuracy and prediction with recurrent neural network (RNN) pipelines. The algorithm (**Fig.3H**) starts with the data acquisition from eight different sensors, followed by real-time preprocessing to filter out the unwanted noise and remove the baseline shift. A feature extraction approach then recognizes the gait patterns from the preprocessed signals (Fig.S22) and passes them to the GRU model, which eventually predicts gait attribution. A principal component analysis (PCA) shows the variation of all 924 signal samples collected from on-body tests (**Fig.3I-J**). The distribution of the three most significant principal components (PCs) demonstrates the differentiable distribution of the samples due to the compelling clustering (**Fig.3I**). The weight of the select feature was elaborated on each PC, distinctly demonstrating the similarity within the data from each foot and the disparity between the two feet (**Fig.3J**). Piezoelectrically actuated OXene ionogels, featuring the high-sensitive data stream, enable the augmented classification over MXene-based counterparts (Fig.S23). The training curve shows an evident convergence based on the tendency of the loss towards 0 and the accuracy towards 1 at around the 100$^{th}$ epoch (**Fig.3K**). The corresponding confusion matrix shows that 11 of the 13 labels have a prediction accuracy of over 85% (**Fig.3L**). The remaining three labels, despite being lower than 85%, still show the maximum value at the correct prediction among all possible ones. The predicted score of all 924 samples is highlighted in the sequence of the true label from 1 through 13 (**Fig.3M**), confirming that the ML-assisted pressure array based on OXene-filled ionogels can offer a precise gait prognosis.

Piezoelectric OXene is capable of highly sensitive pressure detection, not only augmenting the soft elastomer, but also promising to interact with other artificial haptic technologies involving pressure interfaces, such as robotic haptics, automatic pilot and drive, and virtual reality, achieving reliable control and manipulation of an artificial robot.

### OXene-mediated Reconfigurable Logic in Flexible Field Effect Transistor

Reconfigurable logic circuits, enabled by Field Effect Transistor (FET) technology, offer denser, faster, and more energy-efficient alternatives to conventional transistor circuits(*31*). Facile oxygen treatment of OXene modifies its interaction with the upper semiconductor layer, enabling the creation of reconfigurable computational substrates that can change logical functions.

To investigate the emergent logic gates, we correspondingly fabricated the 5 x 5 active transistor matrix with side gate architecture, including (i) polydimethylsiloxane (PDMS) as substrate and encapsulation layers; (ii) large-area, uniform metallic OXene/MXene traces as the source-drain-gate (SDG) layer; (iii) a blend of poly(3-hexylthiophene-2,5-diyl) nanofibrils and PDMS (P3HT-NFs/PDMS) as the semiconductor layer; (iv) PVDF-HFP mixed with [EMIM][TFSI] as the dielectric layer (**Fig.4A**). While we showcase a 5 × 5 array, it is noted that increasing the density of sensing nodes is feasible by reducing both the linewidth and the area of each sensing cell. Each pixel operates as a p-channel metal–oxide–semiconductor (PMOS) transistor, with gate electrodes of each column linked to a word line ($V_{WL}$) and source electrodes of each row connected to a bit line ($V_{BL}$). This transistor array endured various mechanical deformations, including stretching, poking, and crumpling, without sustaining physical damage (**Fig.4B** and Video.S4), promising to establish conformable contact with tissue.

OXene and MXene characterize distinct transfer curves with differing on/off ratios, enabling the formation of reconfigurable computational substrates representing "0 and 1". The transfer-curve peak of OXene shows an on/off ratio exceeding $10^5$, illustrating ideal charge-carrier mobility, a $10^4$-fold improvement over MXene with the identical semiconductor and dielectric layers (**Fig.4C**). Besides, the switching ratio of OXene transistors varies from $10^3$ to $10^5$ based on the duration of oxygen treatment, eventually reaching saturation (**Fig.4D**).

OXene-mediated FET displays an impressive p-type output curve across varying gate voltages ($V_{GS}$) (**Fig.4E**). The output curve indicates nonlinear behavior in the pre-$V_{DS}$ region and saturation behavior in the post-$V_{DS}$ region. Nonlinearity in the pre-$V_{DS}$ region stems from Schottky contact, comprising: (i) the OXene's internal and (ii) OXene-semiconductor layers. Built-in field bending at Schottky interface suppresses hole-electron recombination, enabling the output curve to achieve a record-high drain current ($0.23 \pm 0.02$ mA) amongst reported P-type MXene-based FET (the best previous report was 0.094 mA to our knowledge(*32*), while three orders of magnitude above unoxidized transistors. The derived $TiO_2$ acts as an N-type semiconductor, forming a P-N heterojunction with P-type P3HT, facilitating directional electron transfer analogous to P-doping in P3HT. Conversely, pristine MXene, with its electronegative surface, traps electropositive holes in P3HT, creating a depletion layer that hinders carrier movement in P3HT, akin to N-doping in p-type semiconductors (**Fig.4F**).

To demonstrate reconfigurable logic gates, we devised NAND and NOR circuits based on OXene/MXene coexistence traces through selective oxidation *via* shadow masks, integrating with P-type P3HT. For NAND circuits, two drives ($T_D$) are aligned in parallel, whereas for NOR, they are arranged in series. These configurations randomly mix OXene and/or MXene elements to produce four distinct SDG designs, each then serially connected to an OXene-based load transistor ($T_L$). The reconfigurable logic gates introduce a new programmable dimension, yielding 16 functional outcomes. Specifically, in the NAND configuration, the logic state "1" occurs in eight of the combinations, equivalent to the frequency of the logic state "0" and thus maintaining a 1:1 output distribution ratio (**Fig.4G**). This departs from the standard 1:3 ratio, improving signal integrity by reducing bias and timing skew. This equilibrium simplifies digital designs and lowers component needs, beneficial for fast and intricate computational systems(*33*). In the NOR configuration, a unique 1:15 "1 and 0" ratio enhances gate specificity and selectivity, promising more nuanced, condition-specific logic operations (**Fig.4H**). Remarkably, this reconfigurability stems from dual symmetry breaking, altering physical properties of the original SDG layer, rather than stacking functional layers. This flat architecture promises customizable logic functions within a compact, energy-efficient framework.

We further developed a fully flexible pressure sensing matrix to demonstrate the viability of the reconfigurable logic gates in bioelectronics (**Fig.4I**). Our approach involves: (i) constructing a 5 by 5 active pressure sensing matrix with either OXene or MXene mediated SDG layer, resulting in the distinct voltage transfer characteristics, (ii) forming the epitaxial drain by using OXene-infused ionogels interconnected with conductive via-holes, and (iii) aligning ionogels with OXene electrode arrays on flexible substrates, supplying the drain voltage. This flexible architecture allows for uniform force distribution, even under 10% biaxial deformation, facilitating the conformal interfaces deployed on soft tissues (Fig.S24). We applied stresses both perpendicular and parallel to the channel length direction, mimicking the tactile sensation of skin (**Fig.4J**). OXene-mediated FET arrays enable recognition of the geometric distribution of tactile sensations, over 4 orders of magnitude higher in sensitivity than that of MXene, thus allowing the former termed as the logic state "1", while the latter as "0" (**Fig.4K**).

**Wireless monitoring and Spatiotemporal mapping with OXene-integrated implants for ML-enabled clinical prediction**

Wireless bioelectronics facilitate full implantation into the body to eliminate the need for percutaneous hardware, thereby minimizing the risk of device-associated infections and dislodgement(*34, 35*). Development of untethered OXene-integrated implants, decisive for its practical clinical applications.

We present a flexible, biocompatibility, and leadless OXene cardiac patch that operates in a battery-free fashion and permits externalized control through resonant inductive coupling (**Fig.5A**). In vivo validation using a porcine whole-heart model, closely mirroring human physiology, underscores its clinical applicability due to the bearing high resemblance of porcine cardiovascular system to that of humans (Fig.S25a)(*36*). The fabrication approach consists of (i) upper OXene bioelectrodes for pacing, (ii) lower OXene ionogels for visualizing myocardial cycles (**Fig.5B**), and (iii) extended passive electromagnetic resonance sensing (PEMRS) and magnetic resonance wireless power transfer (MRWPT) coil embedded in porcine dermis for power and signal transmission (Fig.S25b-d).

This architecture integrates distributed wireless sensor and actuator networks to monitor and modulate cardiac physiological activity. The stimulator can generate an electrical field approaching 3 V/mm near the inner electrode with a 5 V DC input, alleviating slow ventricular depolarization and conduction-system dysfunction during acute myocardial ischemia (Fig.S25e)(*37*). Fully flexible mechanical layouts ensure a seamless interface with the epicardium and concurrently deform with a beating heart (**Fig.5C** and Fig.S25f).

OXene-integrated wireless implants maintain stable function throughout the postoperative period. OXene-filled ionogel exhibits a monotonic correlation between pressure-induced capacitive change and resonant frequency (**Fig.5D**), that is capable of being qualified via a Vector Network Analyzer (VNA). Clinically, a VNA probe aligns with an under-chest soft inductor coil, facilitating wireless signal transmission a week post-MI surgery (**Fig.5E**). It entitles real-time heartbeat interval monitoring and captures the dynamic amplitude of myocardial contractions (**Fig.5F** and video.S5).

The scalability of OXene promises spatiotemporal resolved mapping via active-matrix multiplexing. We devise a 16-channel synchronous pressure patch mounted on the myocardium (**Fig.5G**), leveraging OXene's optimized interfacial impedance and piezoelectric effects for conductive plates and sensing pixels (Fig.S26a). Signal processing was achieved by sequentially filtering out ambient and motion-related noise (Fig.S26b), then interpolating to illustrate the myocardial contractile propagation within each time frame (**Fig.5H**). The 3D stacked mapping (**Fig.5I** and Fig.S26c-d) vividly contrasts pressure variations between sinus rhythm and MI states, assessing alterations in cardiac contraction amplitude and rhythm due to myocardial damage and fibrosis. This approach allows for the precise identification of regions with cyclic contraction, especially from local myocardial ischemia post-MI (Fig.S26e-f and video.S6).

The spatiotemporal mapping facilitates the classification and prediction of MI stages using a convolutional neural network-gated recurrent unit (CNN-GRU) architecture. This network, illustrated in **Fig.5J**, begins with preprocessing that includes peak location and sample extraction to register data slices for subsequent analysis (Fig.S27). Thereafter, dual convolution networks serve to encode both spatial and temporal dimensions. Eventually, the GRU network is used to predict the MI status based on the processed data. The CNN-GRU model demonstrates promising convergence, achieving a training accuracy of 89.4% with a loss of 0.229, and a testing accuracy of 84.5% with a loss of 0.351 (**Fig.5K**). The algorithm distinctly differentiates between sinus rhythm and various stages of myocardial infarction—early, middle, and late (**Fig.5L**). The spatiotemporally resolved OXene-integrated signal is capable of predicting adverse clinical events using the ML pipeline, while enabling detailed patient screening and progression monitoring.

OXene and its constituent devices feature superior biocompatibility over other common bioelectrical traces (e.g., Au and PEDOT:PSS). This is supported by cytotoxicity assays including the cell Counting Kit-8 (CCK8) (Fig.S28b) and Live/Dead assays (Fig.S28c) with neonatal rat cardiomyocytes (NRCMs), showing minimal impact on cardiomyocyte viability over 48 hours of co-culturing. To assess the level of inflammation in the heart following open-chest surgery and OXene implantation, we used immunohistochemistry to assess short-term (CD68, CD11b) and long-term (CD3) inflammation markers (Fig.S29). Quantification of OXene implants shows lower levels of inflammation than others after implantation, indicating that the OXene and associated implantation surgery do not provoke an inflammatory response. Histological examination of tissue adjacent to the OXene device up to two weeks post-implantation employed Masson's trichrome staining to identify fibrotic areas (Fig.S30). The quantification of fibrotic tissue thickness presents OXene with significantly lower fibrosis than Au and PEDOT:PSS (Fig.S30b). The results indicate normal tissue structure, with marginal fibrosis confined to the tissue's outer layers at the device contact point. Overall, results show that OXene induces minimal cytotoxicity, inflammation, and fibrosis *in vivo*.

**Conclusion**

We harnessed symmetry engineering to augment 2D bioelectronic interfaces. By coupling orbital and inverse symmetric breaking, OXene achieves improved electronic-tissue impedance and Schottky-induced piezoelectric effects. These features enable OXene's applicability to different domains, including bioelectrode arrays, gait analysis, and wireless cardiac patches, entitling multiplexed, high-fidelity and spatiotemporally resolved physiological recordings in both rodent and porcine models. Integration with machine learning pipelines demonstrates accurate physiological state classification and prediction. Simultaneously the interface-polarized OXene can be integrated with MXene to achieve distinctive reconfiguration computational substrates in active matrix, promising a flattened programmable dimension for dense, swift, and energy-efficient bioelectronic frameworks. On a broader way, the symmetry modulation allows for numerous augmented 2D bioelectronics interfaces. This design principle may offer a groundbreaking attainment upon existing 2D bioelectronics, especially in clinical practice.

**Acknowledgment:** This work was supported by the start-up funds from University of North Carolina at Chapel Hill and the fund from National Science Foundation (award # ECCS-2139659). Research reported in this publication was also supported by the National Institute of Biomedical Imaging and Bioengineering at the National Institutes of Health under award number 1R01EB034332-01.

**Author contributions:** Yizhang Wu, Yihan Liu, and Yuan Li equally contribute to this work. Y.W. and W.B. conceived the ideas and designed the research. Y.W., Y.L., S.X., Y.W., Z.G., A.Z., and G.Y., fabricated and characterized the samples. Y.L., Y.W., Z.Z., and S.X. performed mechanical modelling and simulation. L.Y., D.Z., K.H., and K.C performed the ex-vivo and in-vivo studies. Y.L., S.X., Y.W., performed the data analysis. Y.W., Y.L., Z.W., and W.B. wrote the manuscript with input from all authors.

**Data and materials availability:** All data needed to evaluate the conclusions in the manuscript are present in the manuscript and/or the Supplementary Materials.

**Supplementary Materials**

**This PDF file includes:**

Materials and Methods

Note S1 to S5

Table S1 to S5

Figs. S1 to S30

Legends for Movies S1 to S6

References

**Other supplementary material:**

Supplementary movie S1: Dynamic demonstration of systolic and diastolic stress distribution upon the OXene-infused 3 x 3 bioelectrode array.

Supplementary movie S2: Video of continuous spatial mapping of myocardial electrophysiological signals in real-time during both sinus rhythm and an MI rhythm.

Supplementary movie S3: Video of wireless gait analyzer collecting synchronized 8-channel data during actual walking.

Supplementary movie S4: Dynamic demonstration of systolic and diastolic stress distribution upon the wireless OXene-integrated patch.

Supplementary movie S5: Video of real-time data acquisition two weeks after the implantation of a wireless stress sensor into a porcine heart.

Supplementary movie S6: Video of continuous spatial mapping of myocardial contraction in real-time during both sinus rhythm and an MI model.

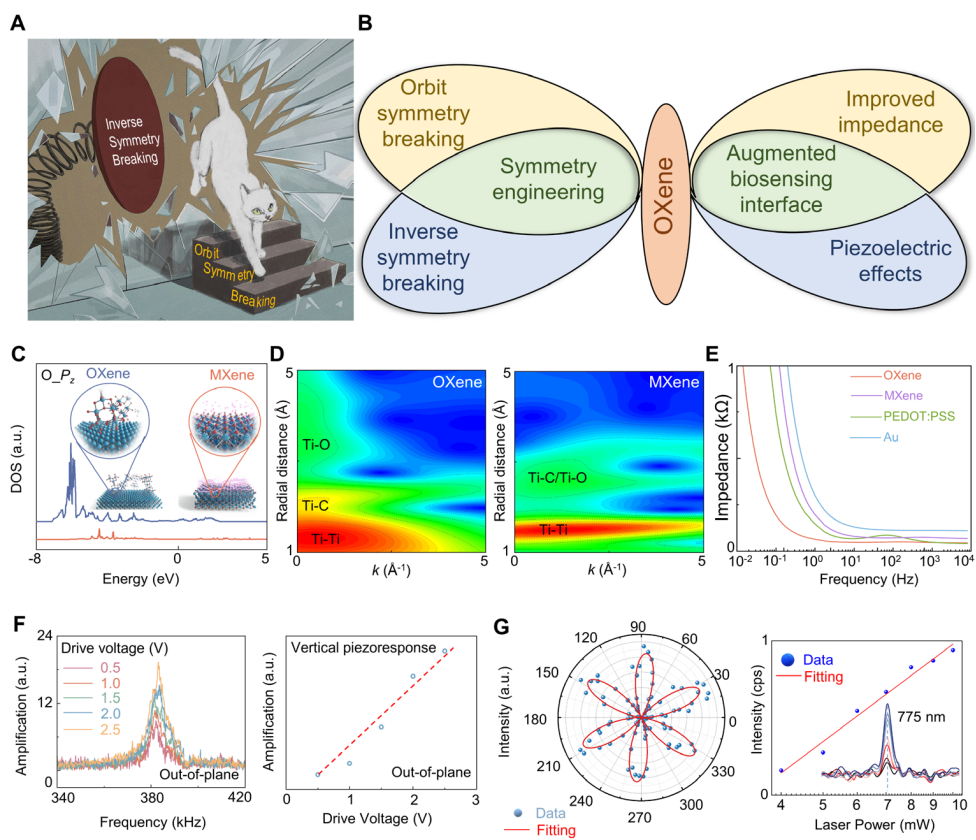

**Figure 1 Dual symmetry engineering in 2D bioelectronics implements the augmented sensing interface. A** Metaphorical illustration of OXene. The white cat represents various 2D bioelectronics, enabling breakthrough performance with symmetry engineering. The stairs symbolize that orbital symmetry breaking enables out-of-plane conductive pathway; the disc symbolizes that inversion symmetry breaking occurs at the Schottky heterojunction interface, and the springs symbolize the piezoelectric effect induced thereby. **B** Schematic illustration of the principle of dual symmetric breaking in OXene, featuring improved impedance and Schottky interface-induced piezoelectric effects, which originate from orbital and inverse symmetric breaking, respectively. **C** PDOS of O atoms projected on the $P_z$ orbital in OXene and MXene ($Ti_3C_2O_x$). PDOS, projected density of states. Inset: Schematic atomic structures of OXene and MXene with corresponding localized enlargements. **D** Morlet wavelet transforms of Ti K-edge $k^2$-weighted EXAFS spectra for OXene (left) in comparison to that of lyophilized monolayered MXene (right). EXAFS, extended X-ray absorption fine structure spectroscopy. **E** Electrochemical impedance spectra (EIS) measured in 5 mM $K_4[Fe(CN)_6]$/$K_3[Fe(CN)_6]$ (1:1) solution for the OXene electrode in comparison to that of MXene, Au, and poly(3,4-ethylenedioxythiophene) polystyrene sulfonate (PEDOT:PSS) electrodes. **F** Vertical resonance peaks (left) of OXene at approximately 385 kHz with gradient voltages, showing linearity of the vertical resonance peak (right) intensity concerning the drive voltage (ranged 0.5 V~2.5 V, 0.5 V intervals). **G** Polar plot (left) and excitation power plot (right, log-log scale) of the SH intensity from monolayer OXene as a function of the crystal's azimuthal angle θ. Inset: SHG spectra of the various excitation powers. The SH component is measured perpendicular to the excitation polarization. SHG, second harmonic generation.

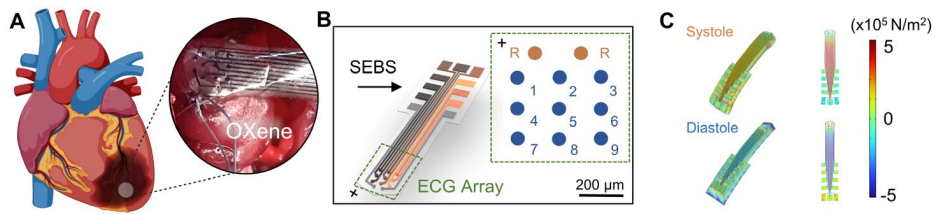
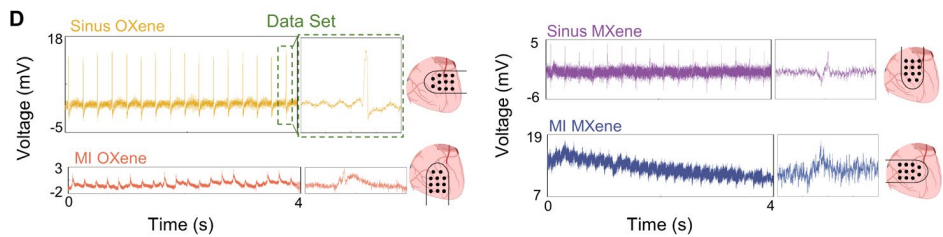
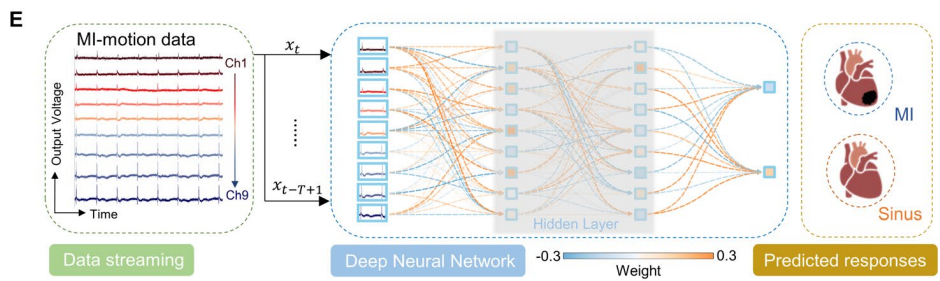
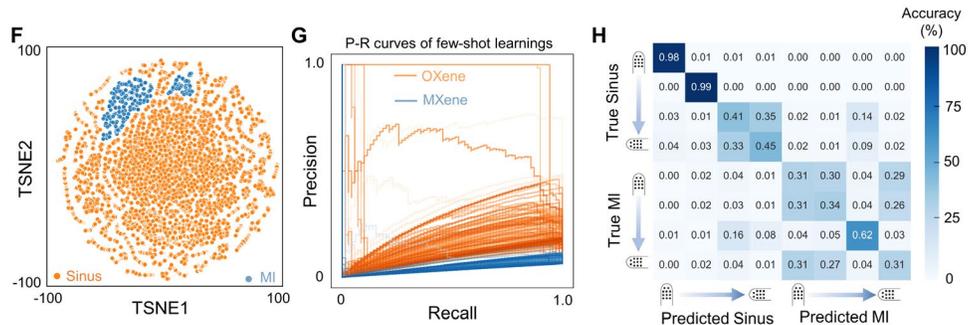
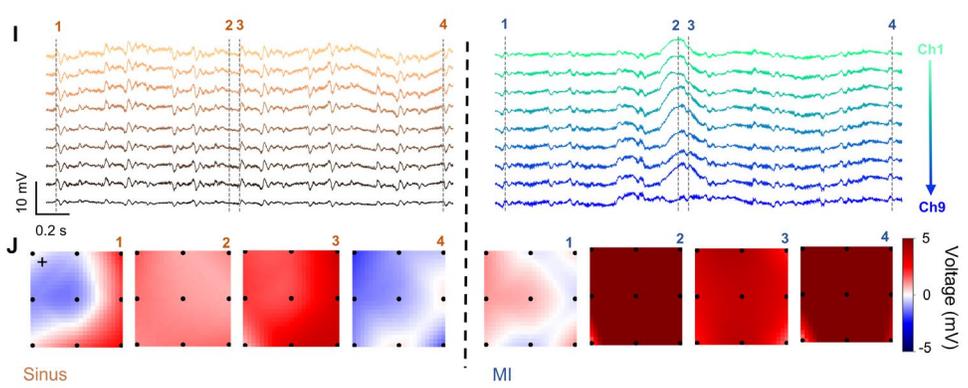

**Figure 2 Orbit symmetric breaking in OXene enhances performance of bioelectric interface and enables physiological prediction powered by machine learning (ML)**. **A** Optical image of the OXene-based multichannel electrodes mounted to the rat epicardium. Coronary artery ligation was performed in the left ventricle to simulate acute myocardial infarction (MI). **B** Schematic illustration of a 3 x 3 OXene electrode array, illustrating the size and serial number of each unit. Elastomeric styrene-ethylene-butadiene-styrene (SEBS) substrate enables conformal and stretchable attachment on the epicardium. **C** Simulation results for the distribution of systole and diastole stresses at the interface between the electrode array and underlying cardiac tissue during biaxial stretching. **D** Evaluation of unfiltered raw data obtained from rat myocardium w/wo MI upon OXene and MXene electrodes featuring various spatial placements of electrodes on the epicardium. **E** Schematic illustration of the ML architecture for data preprocessing, deep neural network, and predicated responses, anticipating an MI or a sinus electrophysiological status. **F** *t*-distributed stochastic neighbor embedding (TSNE) plot from the OXene dataset, visually showing feature separation (sinus and MI) in a two-dimensional space. **G** Precision–recall curve of few-short learnings classified from OXene and MXene interface. **H** Confusion matrix displaying the spatial distribution accuracy for predicting either Sinus or MI in OXene dataset. **I** Segments of the recorded ECG signal from the OXene electrode array on a living rat heart with sinus rhythm (left) and MI (right). The data are displayed following the spatial arrangement of the electrode array (**B**). **J** Instantaneous electrophysiology mapping using OXene electrodes corresponds to sinus rhythm (left) and MI molding (right). The mapping frames selected here correspond to the respective four synchronous points highlighted by the vertical dashed lines in (**I**).

**Figure 3 Inverse symmetric breaking in OXene serving as ionogel fillers augments pressure sensitivity. A** Schematic diagram of the sensor layout. The signaling input was achieved by OXene-filled pressure sensing ionogels mounted on interdigital electrodes. Targeted regions for pressure sensing feature the anterior transverse arch, medial longitudinal arch, lateral longitudinal arch, and heel, discriminating individual behavior. **B** Schematic illustration showing a molecular network of the OXene/MXene-filled ionogels. 1-ethyl-3-methylimidazolium ethyl sulfate (EMIES) coupling with poly(acrylic acid) (PAA) polymers and Poly (acrylamide) (PAAm) polymers form a stiff but brittle polymer network. The electronegative OXene/MXene fillers are uniformly distributed in the matrix of the ionogel via hydrogen bonding with an electrostatic self-assembly mechanism. The ingredients of the ionogel and the corresponding molecular structure formula are listed on the right. **C** Schematic illustration of OXene-filled ionogels. Left: The electronegative fillers anchored in the planar domains due to the negatively charged surface and the positively charged ion chains by hydrogen bonds. Right: Interlaminar anatase $TiO_2$ bridging the electron transfer channel among the confined in-plane nanoflakes. **D** Schematic illustration of a Schottky junction constructed from metallic MXene and P-type anatase $TiO_2$ showing a potential modulation in the depletion region with a built-in field. The piezoelectric effect is induced by the Schottky-interface inverse symmetry breaking, entitling the charge redistribution when the Schottky junction is subjected to a tensile/compressive force. **E** Real-time measurements of relative voltage outputs of OXene/MXene-filled ionogel (left) and localized segments (right) with cyclic strains, indicating a superior strain sensitivity from OXene fillers. **F** Optical image of an OXene-filled gait sensor settled on a human subject, consisting of a four-ionogels group (signaling interface) and a printed circuit board (main functionalities: signal amplification, filtering, and Bluetooth signaling transmission). Scale bar, 5 cm. **G** Spectrum of wireless gait sensing data featuring multichannel time synchronicity gathered from one subject. All signals were calibrated and normalized to against patch variations and any motion artifacts above a moderate level. **H** Data processing protocol. From left to right: i) Data gathered from subjects with different foot sizes, using OXene- and MXene-filled gait sensors, respectively. ii) Data preprocessing and feature extraction from the collected data from gait sensors. The step peaks from both left and right feet are extracted from the denoised signal. iii) Neural-network-based gait type classification algorithm, highlighting the layer of gated recurrent units (GRU). iv) Prediction generation based on the classification algorithm that yields the predicted gait type among all the input types. **I** Scatter plot showing the distribution of the three leading principal components (PCs) in the principal component analysis (PCA) of all the extracted features on 16 selected features. **J** Proportion of all 16 selected features in the first two PCs. **K** Training curve of the classification algorithm. Solid lines show the curves for training data, while the dashed lines show the curves for test data. Blue lines: Loss curves. Orange lines: Accuracy curves. **L** Confusion matrix of the model. Y-axis: true label. X-axis: predicted label. **M** Prediction score for all samples (x-axis). On y-axis, it shows the scores of each sample on each prediction label (Subject 1 through 13, from bottom to top). The topmost plot shows the overall prediction after calculating the max value of the prediction scores.

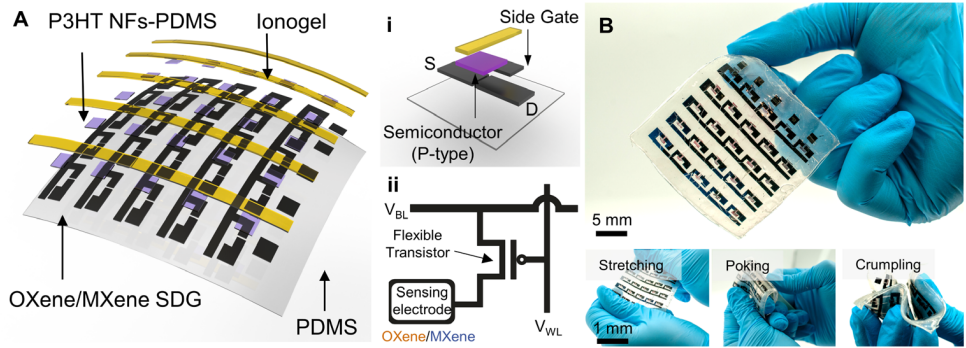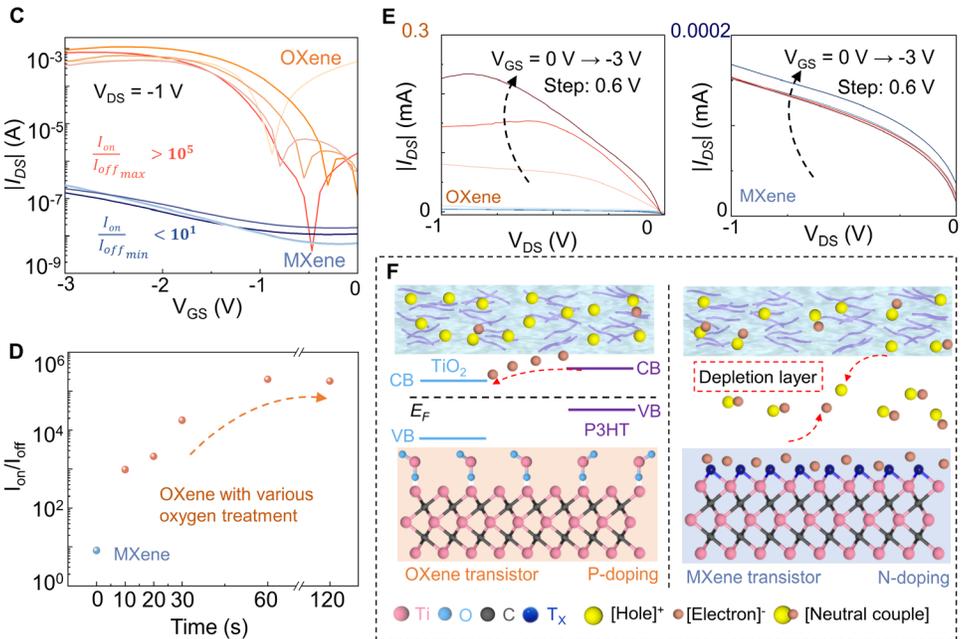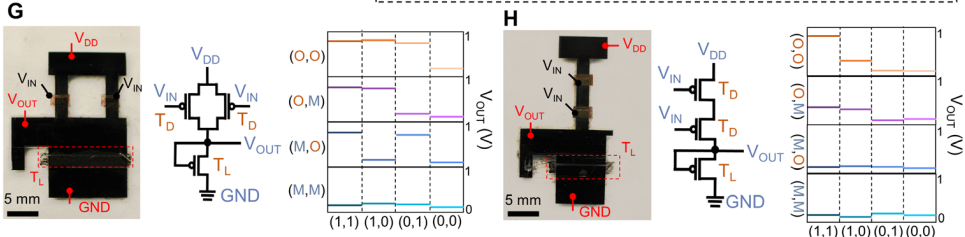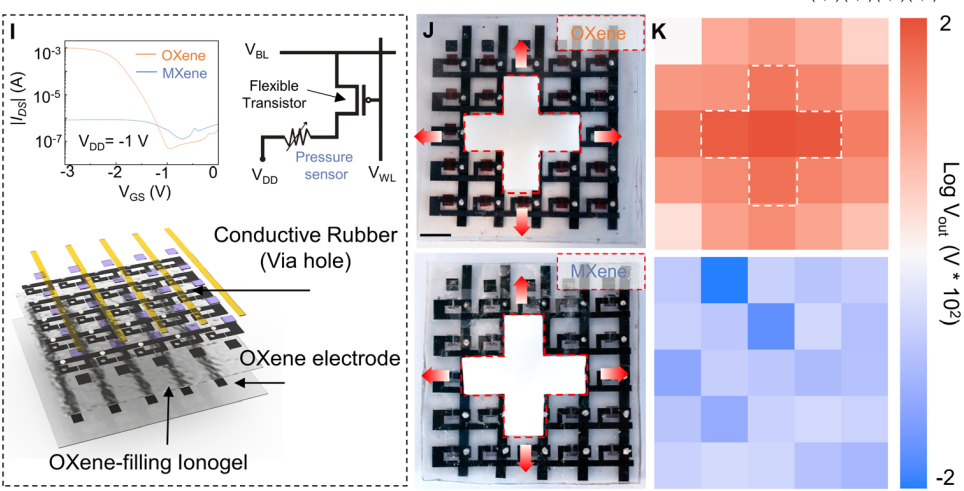

**Figure 4 Symmetry engineering in OXene enables reconfigurable "0" and "1" logic interfaces as SDG layers in flexible transistor arrays. A** An exploded schematic view of the 5 x 5 transistor array with OXene/MXene SDG layer. SDG, Source-Drain-Gate layer. The constituent components include flexible Polydimethylsiloxane (PDMS, 1:10) substrates, OXene or MXene traces serve as transistor interconnects, with poly(3-hexylthiophene-2,5-diyl) nanofibrils-PDMS (labeled as P3HT NFs-PDMS) as the semiconductor, and ionogel as the flexible dielectric layer. (i) Schematic exploded view of the architecture of a transistor node. (ii) The circuit diagram for a single sensory transistor in the active matrix. $V_{BL}$, bit line voltage; $V_{WL}$, word line voltage. **B** Optical image of transistor array under various mechanical deformations. **C** Representative transfer characteristics of the OXene/MXene transistors with a p-type P3HT NFs-PDMS semiconductor layer. The gate current ranged around $10^{-10} \sim 10^{-11}$ A is omitted. Distinct OXene curves are attributed to different oxygen plasma treatment periods (30 cc min$^{-1}$, power 200W 40 KHz, ranging from 10 s to 120 s). $I_{DS}$, drain–source current; $V_{DS}$, drain–source voltage; $V_{GS}$, gate–source voltage. **D** Switching current ratios ($I_{on}/I_{off}$) of OXene transistors under various oxygen treatment periods, in comparison to MXene transistors without oxygen treatment. **E** Comparison of output characteristics of OXene (oxygen plasma treatment: 60 s) and MXene transistors. **F** Schematic illustration of the principle of OXene (left) and MXene (right) SDG as a reconfigurable computational substrate. Profile architecture on OXene transistor illustrates the electron transfer from CB of P3HT to CB of $TiO_2$, featuring further P-doping in P-type P3HT. Upon MXene transistor, the negatively charged MXene surface anchors the holes in P3HT to form the depletion layer, corresponding to execute N-doping in P-type P3HT. CB, conduction band; VB, valence band; $E_F$, Fermi energy level. **G-H** Logic gates constructed from OXene and MXene as SDG layer, P3HT NFs-PDMS as semiconductor layer, and ionogels as dielectric layer. Optical image (left), circuit diagram (medium), and output characteristics (right) of the (**G**) NAND gate and (**H**) NOR gate under $V_{DD}$ of 1 V. The input voltages ($V_{IN}$) of −3 or 0 V represent logic states 1 or 0, respectively. Selective oxidation of S-D region on two $T_D$ is operated by shadow mask. $T_D$, driver transistor; $T_L$, loading transistor. **I** Schematic exploded view of active matrix for pressure sensing. Differing from **A**, the key components OXene-filled ionogel, OXene bottom electrode, and conductive via holes implement the epitaxial Drain and pressure-sensitive signaling transmission. Inset: (left) Transfer characteristics of one pressure-sensitive pixel in the active matrix with OXene or MXene SDG layer; (right) The circuit diagram of one pressure-sensitive pixel. **J** Optical image of the pressure-sensitive active matrix featuring OXene (upper) and MXene (down) SDG layer in contact with the identical block. Scale bar: 5 mm. **K** Corresponding distributive voltage mapping result gathering from OXene (upper) and MXene (down) transistor array. The data are plotted as the means with n = 5 per unit corresponding to SDs (standard deviations). Note: Scalebar optimizes data visualization through logarithmic axes while highlighting data differences between OXene and MXene active matrix.

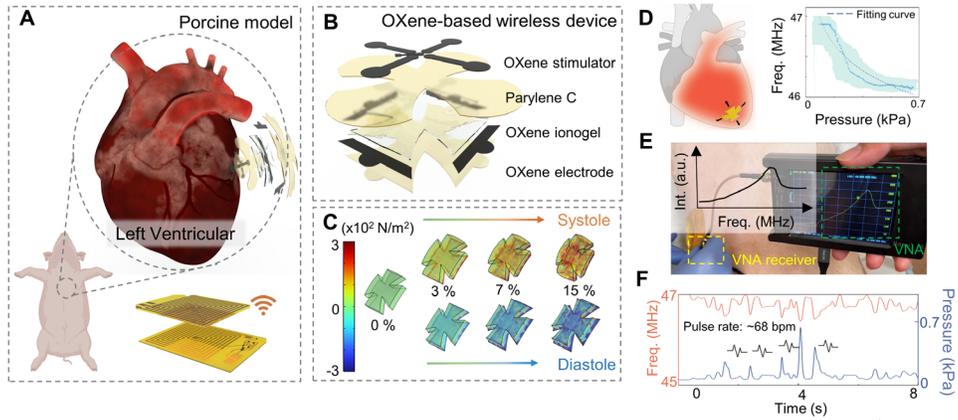
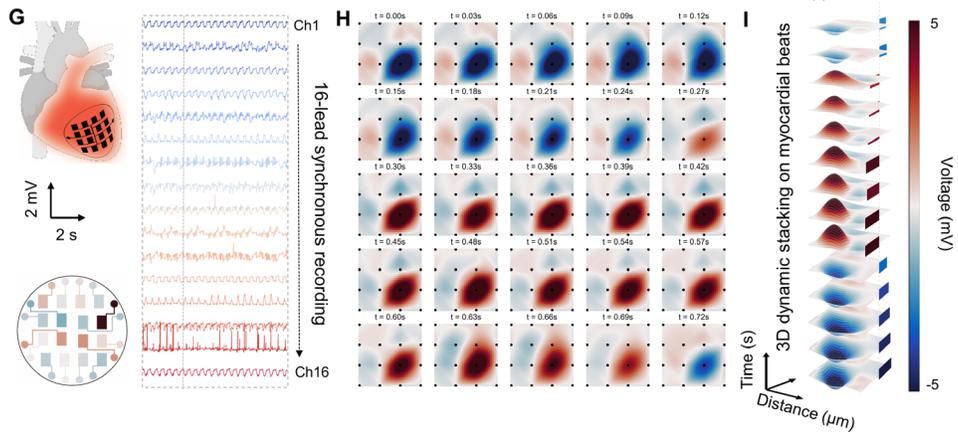
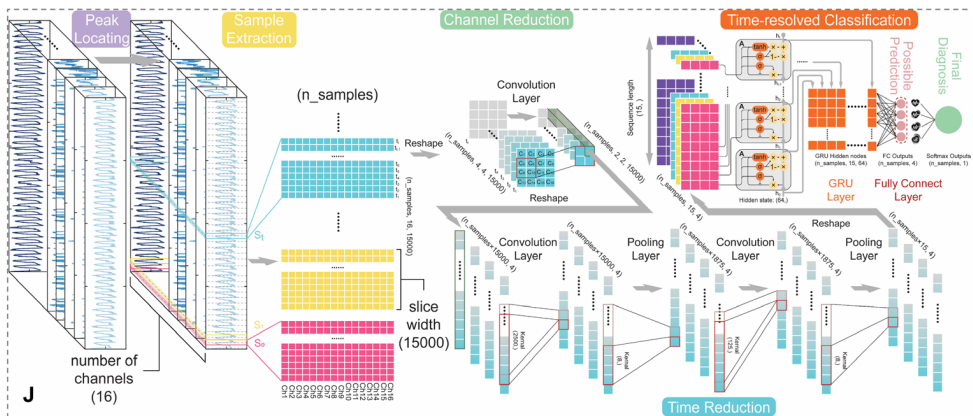
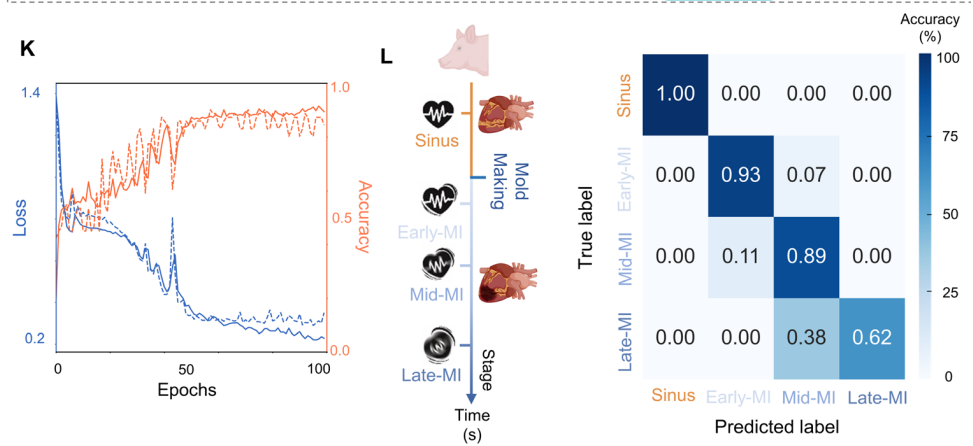

**Figure 5 Symmetry engineering in OXene realizes real-time, spatiotemporally resolved, wireless bioelectronic interfaces, featuring ML-powered physiological response prediction. A** Schematic illustration of an OXene-based wireless device deployed onto a porcine heart near the left ventricle. Inset: Radiofrequency (RF) coil for electrotherapy and RF coil for strain sensing. Both RF coils are placed subcutaneously for optimized signal transmission. **B** Exploded schematic illustration of a wireless OXene patch integrated with a pressure sensor and an electrical stimulator. The upper electrical stimulator consists of four surrounding electrodes and a central reference, all based on OXene traces, encapsulated by parylene C via shadow masks. The pressure sensor is composed of two electrodes based on OXene traces, OXene-filled ionogels, and the parylene C encapsulation layer. **C** Simulated strain distribution using 3D finite element analysis (FEA) to show the mechanical performance of the pressure module under myocardium systole and diastole. Convergence of meshes was tested to ensure computational accuracy. 5 % initial deformation was set considering the device size and conformal adherence to the epicardial surface. **D** Measured calibration curve between signal frequency and pressure using an in vitro setup. The setup features a 25 mm x 10 mm x 3 mm (L x W x H) medium, undergoing the time-synchronized stretch curves of OXene-filled ionogels. **E** Optical image of OXene patch undergoing wireless and real-time detection in vivo to highlight contractility and heartbeat intervals upon MI-molding porcine hearts received by the VNA one week after implantation. VNA: vector network analyzer. **F** Measured resonant frequency gathered from in vivo MI model monotonically converted to the pressure applied to the sensor under time-synchronized segments. The wireless OXene patch monitors the long-term process of cardiac tissue remodeling after an episode of MI. **G** Representative real-time synchronous recording of the OXene patch on a porcine heart in the sinus state. Voltage output was recorded by the 16 pressure-sensitive pixels on the OXene patch at a continuous frame of the data stream. The signals were processed through high-pass, low-pass, and bandstop filtering. **H** Interpolated spatial mapping of the detected pressure featuring sinus signals within a cardiac cycle (~0.72 s, Interval = 30 ms). **I** Corresponding 3D dynamic mappings of sinus state during myocardial contractile propagation superimposed by interpolation and normalization. Specifically, a red bar exhibits an average pressure higher than the baseline, while vice versa signifies a lower average pressure. The X-Y plane shows the spatial distribution of each sensing unit. The vertical direction depicts the time stream, with a time interval of 30 ms per frame. **J** Schematic architecture of the deep-learning classification algorithm. The process starts from left to right: peak locating algorithm that locates the temporal locations of heart pulsation peaks; sample extraction algorithm that slices down the signals into small windows; convolution layer that reduces the number of channels and shortens the sequential lengths; GRU layer to analyze the temporal trend of the data. The algorithm yields a score for all four labels (sinus, early-MI, mid-MI, late-MI) and draws a conclusion based on the label with the highest score. **K** Training curves of the deep-learning model. Solid lines show the training data while the dashed lines show the test data. Blue lines: Loss curves. Orange lines: accurate curves. **L** Left: Schematic illustration of four selected data segments, labeling as Sinus, Early-MI, Mid-MI, and Late-MI. Each segment is 10s in length. Right: The confusion matrix of the model performance. Y-axis: true label (Sinus through Late-MI, from top to bottom), X-axis: predicted label (Sinus through Late-MI, from left to right).